\documentclass{article}
\usepackage{spconf,amsmath,graphicx}
\usepackage{xcolor}
\usepackage{comment}
\usepackage{xspace}
\usepackage{multirow}
\usepackage{enumitem}
\usepackage{cite}
\usepackage{hyperref}

\newcommand{\meanpooling}{\texttt{Mean-Pool}\xspace}
\newcommand{\maxpooling}{\texttt{Max-Pool}\xspace}
\newcommand{\gru}{\texttt{GRU}\xspace}
\newcommand{\lstm}{\texttt{LSTM}\xspace}
\newcommand{\accuracy}{\texttt{Acc}\xspace}
\newcommand{\fmacro}{\texttt{MaF1}\xspace}

\newcommand{\auc}{\texttt{AUC}\xspace}
\newcommand{\aucpr}{\texttt{AUCpr}\xspace}

\newcommand{\prima}{\texttt{ADIMA}\xspace}
\newcommand{\wavvec}{\texttt{Wav2Vec2}\xspace}
\newcommand{\vak}{\texttt{Him-4200}\xspace}
\newcommand{\clsril}{\texttt{CLSRIL-23}\xspace}
\newcommand{\vgg}{\texttt{VGG}\xspace}
\newcommand{\xlsr}{\texttt{XLSR-53}\xspace}

\newcommand{\all}{\texttt{All}\xspace}

\usepackage[normalem]{ulem}

\title{ADIMA: Abuse Detection in Multilingual Audio}
\name{Vikram Gupta, Rini Sharon, Ramit Sawhney, Debdoot Mukherjee
}
\address{ShareChat, India \\\{vikramgupta, rinisharon, ramitsawhney, debdoot\}@sharechat.co}

\begin{document}
%\ninept
%
\maketitle
\begin{abstract}
Abusive content detection in spoken text can be addressed by performing Automatic Speech Recognition (ASR) and leveraging advancements in natural language processing. However, ASR models introduce latency and often perform sub-optimally for profane words as they are underrepresented in training corpora and not spoken clearly or completely. Exploration of this problem entirely in the audio domain has largely been limited by the lack of audio datasets. Building on these challenges, we propose \textbf{\prima}, a novel, linguistically diverse, ethically sourced, expert annotated and well-balanced multilingual profanity detection audio dataset comprising of 11,775 audio samples in 10 Indic languages spanning 65 hours and spoken by 6,446 unique users. Through quantitative experiments across monolingual and cross-lingual zero-shot settings, we take the first step in democratizing audio based content moderation in Indic languages and set forth our dataset to pave future work. Dataset and code are available at: \url{https://github.com/ShareChatAI/Adima}

\end{abstract}

\section{Introduction}
Detecting abusive content in online content has gained a lot of attention due to the widespread adoption of social media platforms. Use of profane language, cyber-bullying, racial slur, hate speech etc. are common examples of abusive behaviour demanding robust content moderation algorithms to ensure healthy and safe communication. Majority of the existing work has focused on detecting abusive behaviour in textual data~\cite{Mathew2021HateXplainAB, ousidhoum-etal-2019-multilingual, i2019multilingual, de-gibert-etal-2018-hate, founta2018large, DBLP:journals/corr/abs-2006-08328}. Abusive content detection on images and videos has been accelerated with the contribution of multimedia datasets~\cite{kiela2020hateful,alcantara2020offensive,anand2019customized,gangwar2017pornography,wu2020detection}. However, abusive content detection in audio has been underexplored primarily due to the absence of audio datasets. Profanity detection in audio can also be addressed by transcribing audio into text using automatic speech recognition (ASR) followed by textual search over the transcriptions. However, this requires accurate ASR systems which require large amount of expensive training data, especially in multilingual setups. Moreover, accuracy of ASR on profane words can be low as they are under-represented in the training corpora. Another paradigm is to formulate this as \textit{keyword spotting task} by using a dictionary of audio exemplars of abusive words and then use template matching approach. However, this does not exploit underlying cues and overall context which can be helpful for identifying profanity. Template matching approaches also fail for novel words and continuously updating the dictionary is time-consuming. Moreover, these approaches have high time complexity and require collection of significant number of reference templates that capture the variations in style/accent/dialect and environmental conditions.  Abusive words are usually not spoken clearly and completely, further limiting the effectiveness of \textit{keyword spotting}. 

To tackle these challenges, we contribute a novel and highly diverse multilingual profanity detection audio dataset - \textbf{A}buse detection \textbf{I}n \textbf{M}ultilingual \textbf{A}udio (\prima). \prima contains 11,775 audio recordings from ShareChat chatrooms with a total duration of 65 hours for 10 Indic languages - Hindi (Hi), Bengali (Be), Punjabi (Pu), Haryanvi (Ha), Kannada (Ka), Odia (Od), Bhojpuri (Bh), Gujarati (Gu), Tamil (Ta) and Malyalam (Ma). The dataset is balanced across the languages and has recordings spoken by 6446 different users making it a highly diverse multilingual and multi-user dataset. The recordings have been extracted from real-life conversations and capture natural and in-the-wild conversations. We also formulate a profanity detection task, where the objective is to classify the audio as \textit{Abusive} or \textit{Non-Abusive}. Since the classifier analyzes the complete audio, it is able to effectively leverage the context and underlying audio properties like pitch, intensity, tone and emotion for robust profanity detection. We setup baselines for monolingual and zero-shot cross-lingual setting for encouraging further research in this direction. \prima presents promise in supervised settings such as automatic moderation of live/recorded audio/video content, social media chatrooms etc. to enable safer interactions. In unsupervised settings also, \prima can be used for large-scale pretraining of models for Indic languages. Competitive crosslingual performance also showcases the strength of \prima to address more languages. Our contributions can be summarized as:
%\textcolor{red}{since this is a dataset - you should end with what challenges it poses e.g. is the performance of your methods bad? what kinds of advances in the field will it require to bridge this gap? or other things ... that can tell us why this dataset is interesting}
%\begin{itemize}[noitemsep,topsep=1pt]
\begin{itemize}
    \item We release \prima, a highly diverse, multilingual, expert annotated audio dataset for profanity detection in 10 Indic languages spanning a total of 65 hours comprising 11,775 total samples. 
    \item We introduce \textit{profanity detection task} and report baseline monolingual results capturing nuances of different languages and architectures.
    \item Competitive crosslingual and joint training results exhibit the potential of \prima for profanity detection in other languages and possibility of having a single unified model for all the languages.
\end{itemize}

\vspace{-6mm}
\section{Related Work}
Abuse detection in textual data under multilingual and monolingual settings has received lot of attention from the community~\cite{Mathew2021HateXplainAB,ousidhoum-etal-2019-multilingual, i2019multilingual, bosco2018overview, founta2018large, DBLP:journals/corr/abs-2006-08328}. Video datasets for identifying offensive videos and a weakly annotated dataset of videos from YouTube labelled for profanity detection are contributed by~\cite{alcantara2020offensive} and ~\cite{anand2019customized} respectively. Datasets to identify specific cases such as pornography and child abuse in images and videos~\cite{gangwar2017pornography} and multimodal hateful memes identification using visual and textual features ~\cite{kiela2020hateful} are also present. A YouTube video dataset to identify racist, sexist and normal videos is also contributed by ~\cite{wu2020detection}. While the above mentioned datasets exist for abusive content detection in text, image and video modality, audio modality has been rather under explored. Recently, ~\cite{yousefi2021audio} explored self-attentive networks for toxic language classification. Our dataset \prima is an attempt to reduce this gap. Audio classification is a well studied area and has been accelerated with the presence of large scale datasets~\cite{hershey2017cnn,chen2020vggsound,alcantara2020offensive}. Audio classification has been performed using Gaussian Mixture Models, Support Vector Machines over Mel Frequency Cepstrum Coefficients, Convolutional Neural Network (CNN)~\cite{hershey2017cnn} and Recurrent Neural Networks (RNN)~\cite{xu2018large}. Finetuning and extracting representations from transformer based models~\cite{baevski2020wav2vec} which are trained using unlabelled raw audios has also gained significant interest and we leverage these methods for our task.

\begin{figure*}[h]
\begin{minipage}[b]{0.24\linewidth}
  \centering
  \centerline{\includegraphics[width=\linewidth]{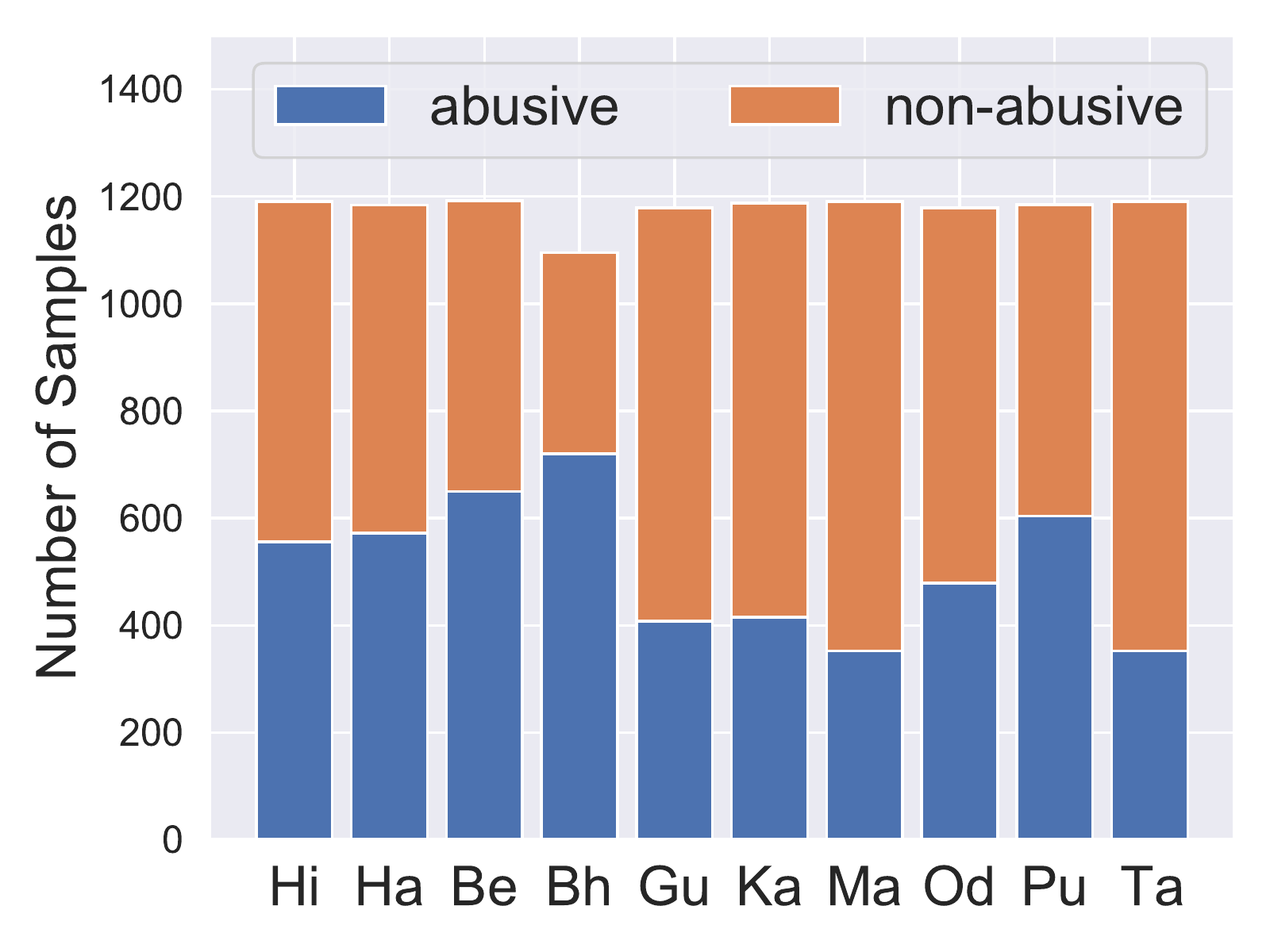}}
  \centerline{(a) Dataset Distribution}
\end{minipage}
\begin{minipage}[b]{.24\linewidth}
  \centering
  \centerline{\includegraphics[width=\linewidth]{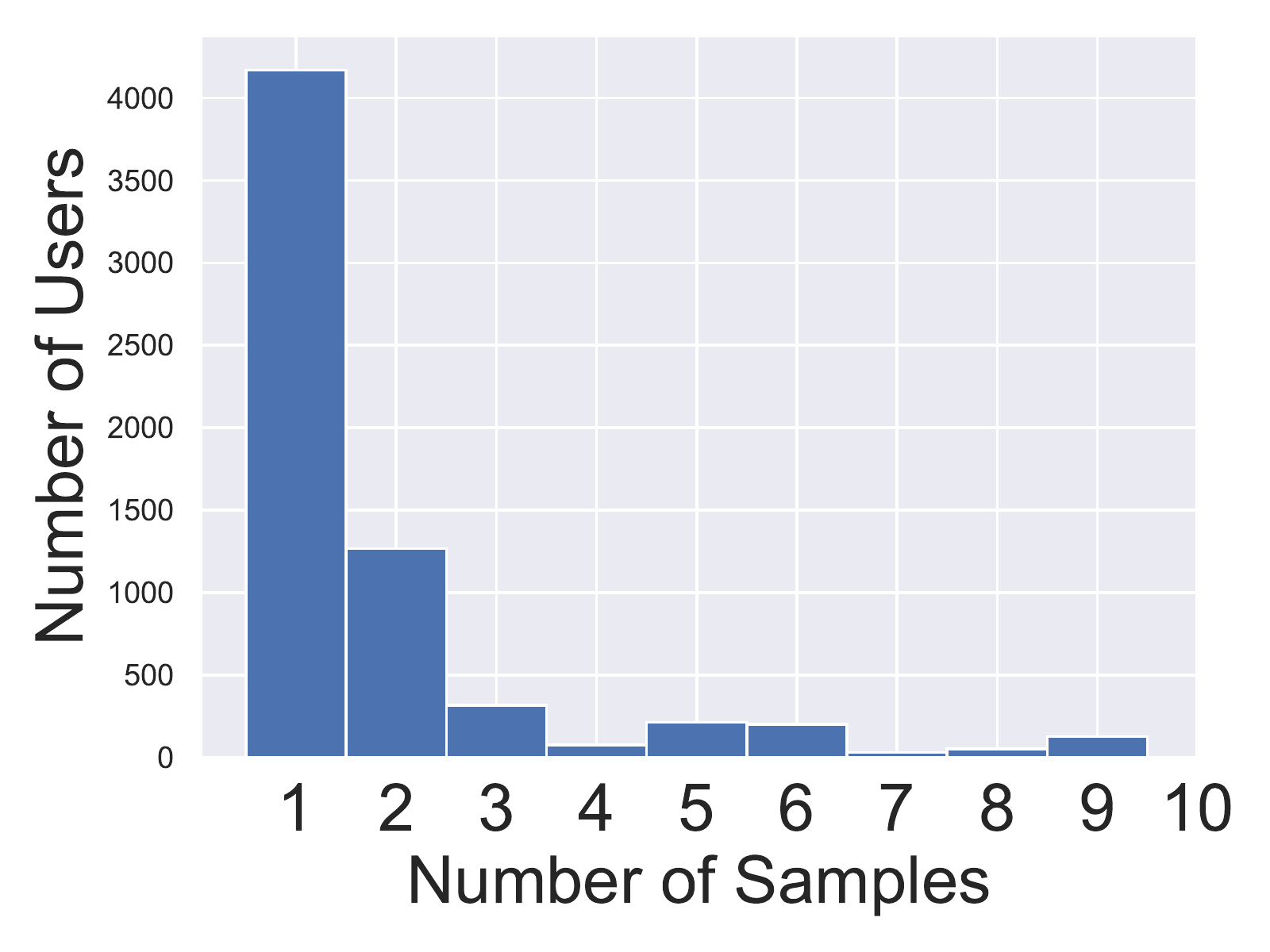}}
  \centerline{(b) User Distribution}
\end{minipage}
\begin{minipage}[b]{0.24\linewidth}
  \centering
  \centerline{\includegraphics[width=\linewidth]{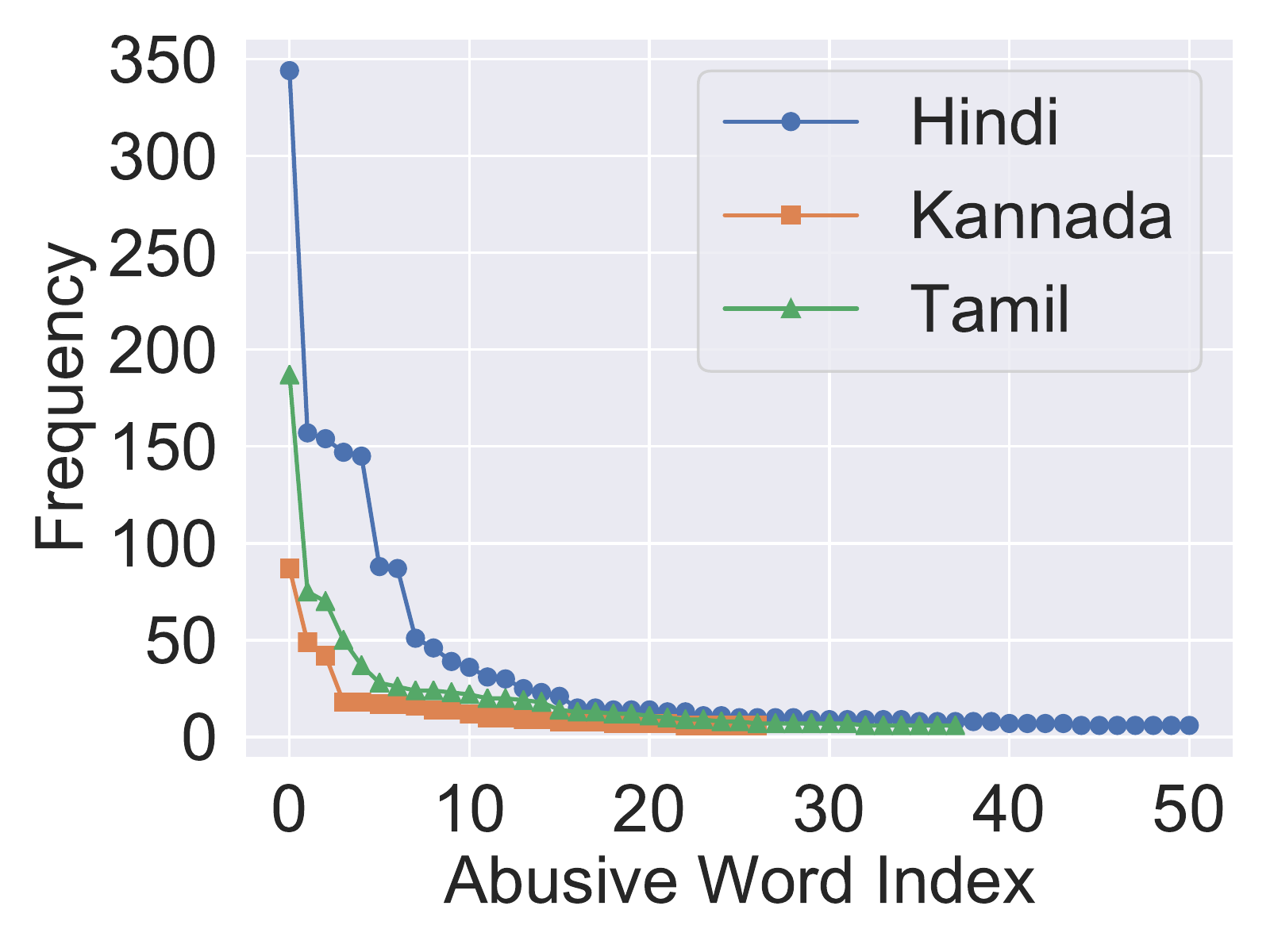}}
  \centerline{(c) Unique abusive words}
\end{minipage}
\begin{minipage}[b]{0.24\linewidth}
  \centering
  \centerline{\includegraphics[width=\linewidth]{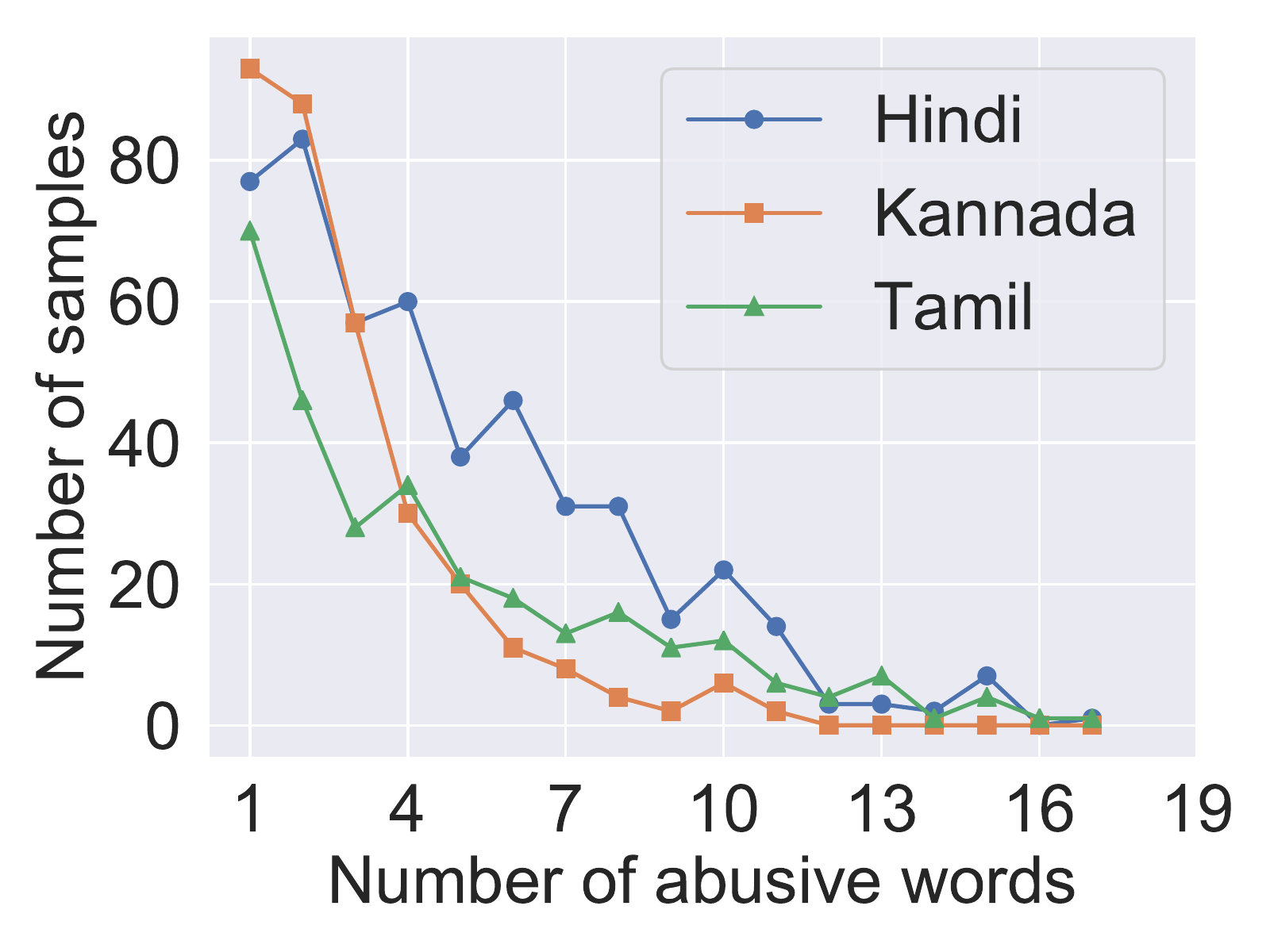}}
  \centerline{(d) Abusive words per recording}
\end{minipage}
\caption{Analyzing \prima across multiple dimensions: (a) Number of samples for \textit{Abusive} and \textit{Non-Abusive} categories for all the languages (b) Unique users for all the languages (c) Frequency of profane words in the dataset (words with more than 5 instances) (d) Frequency distribution of the number of profane words in each sample. (Best viewed in color)}
\label{fig:dataset_stats}
\end{figure*}

\begin{table}[t]
\footnotesize
\caption{\prima statistics across sample distribution, linguistic diversity and audio duration.}
\begin{center}
\begin{tabular}{ lr} 
\textbf{Data Description} & \textbf{Value}\\
\hline
\# Languages & 10 \\

\# Total Samples & 11,775 \\

\# Abusive Samples & 5,108 \\

\# Non-Abusive Samples & 6,667 \\

\# Unique Users & 6,446 \\
\hline

Total Duration & 65 hours \\

Average Duration & 20 $(\pm 3)$ seconds  \\
Min/Max Duration & 5/58 seconds\\
\hline
\end{tabular}
\end{center}

\label{table:dataset_summary}
\end{table}

% \begin{table}[h]
% \begin{center}
% \begin{tabular}{|c|c|c|c|} 
% \hline
%  & Hi & Ka & Ta \\
% \hline
% Hi & NA & NA & NA \\
% \hline
% Ka & NA & NA & NA \\
% \hline
% Ta & NA & NA & NA \\
% \hline
% \end{tabular}
% \end{center}
% \caption{Kappa: Inter-annotator agreement}
% \label{table:kappa}
% \end{table}

\section{\prima Dataset}

\subsection{Data Collection}
Recordings are collected from \textit{public} audio chatrooms of ShareChat\footnote{\url{https://sharechat.com/}}. ShareChat is a leading, Indian social media application supporting over 10 Indic languages with penetration across India. ShareChat's \textit{public} chatrooms are open for anyone to join and informed consent of the users is requested for recording and broadcasting the discussions. The data was collected for a period of 6 months (January-June, 2021) from audio chatrooms pertaining to 10 Indic languages. These chatrooms provide an interactive, audio-only platform for users to speak in regional dialect. To build \prima, we sampled audio from conversations which were reported as abusive by the users in the chatroom. 
Focusing on creating a well-balanced and diverse dataset, we select audio samples across 10 languages in similar proportions from a total of 6,446 users of the ShareChat platform. 
\subsection{Data Annotation}
Specific to each language, an independent set of three annotators per language were employed on a contract basis and fairly compensated to annotate each data sample as abusive or non-abusive. We considered the presence of swear, cuss and abusive words/phrases for annotating an audio as abusive. The abusive words/phrases were catalogued and reviewed to ensure consensus among the three reviewers.
On average, the inter-annotator agreement measured by Cohen's Kappa $\kappa=0.88$ was observed to indicate a high degree of agreement amongst the annotators for each language. The Cohen's Kappa varied from $\kappa=0.77$ to $\kappa=1.0$ across different languages, indicating the variation in annotation complexity and annotator diversity across languages for the same task. In the case of disagreements, the final label was selected based on review by a fourth, expert annotator.
%We note that the agreement is high for these languages which can be attributed to the expertise of the annotators and consensus about profane words. Due to the high annotator agreement, the dataset of other languages was annotated by single annotator per recording. 
Further, we removed low quality recordings after which the final dataset comprises 11,775 audio recordings spanning over 65 hours of audio across 10 languages.

\subsection{Dataset Analysis}
We briefly summarize the key statistics pertaining to \prima in Table 1. The dataset is well balanced (43.38\%) with 6,667 non-abusive and 5,108 abusive recordings. We now analyze \prima across four different dimensions.

\noindent\textbf{Language Distribution}: 
Figure~\ref{fig:dataset_stats}(a) shows the statistics for each language. We note that on average, each language is almost equally represented in \prima. Further, while \prima is well balanced on average, there exist class imbalance-related variations across languages. 

\noindent\textbf{User Distribution}: Figure~\ref{fig:dataset_stats}(b) depicts the frequency distribution of the number of samples in the dataset spoken by each individual user. On average, there are around 1,500 (~23.7\%) users that have spoken more than one audio recording in the dataset, indicating a diverse set of users while also presenting a promising opportunity of potential user profiling and leveraging similarity across the samples spoken by the same user for improved performance.

\noindent\textbf{Vocabulary Distribution}: In Figure~\ref{fig:dataset_stats}(c), we plot the frequency of profane words which occur more than 5 times in the dataset. We note that Hindi has around 50 dominant profane words, followed by Tamil and Kannada with 35 and 25 words. Overall, there are 1059, 820 and 690 unique profane words in Hindi, Tamil and Kannada respectively. The distribution of unique words showcases the diversity of the vocabulary. 

\noindent\textbf{Profanity Distribution}: In Figure~\ref{fig:dataset_stats} (d), we plot the distribution of instances of profane words present in the dataset for three languages. Lot of recordings have lesser than 5 profane words making it challenging to spot the profanity while some recordings are majorly abusive. Each abusive recording contains atleast one profane word. 

\prima is highly diverse across languages, users, vocabulary and density. The recordings are sampled at 16kHz, mono-channel and range from 5-60 seconds with an average duration of 20 seconds. We randomly split the dataset in 70:30 ratio for each language to form the train and test set.

\section{Formulation and Methodology}

\subsection{Problem Formulation} We consider the task of classifying audio recording $x$ into $c$ $\in$ \{\textit{abusive}, \textit{non-abusive}\} categories. We extract features using \vgg~\cite{hershey2017cnn} (pretrained over audio dataset) and \wavvec~\cite{baevski2020wav2vec} (pretrained over speech datasets) as backbones. The features are then aggregated across temporal dimension and are passed through a fully connected classifier for classification. 
\subsection{Feature Representations}
\textbf{\vgg}: Log-mel spectrograms are extracted from raw audios with the window and hop length as 25ms and 10 ms for short-time fourier transform following~\cite{hershey2017cnn}. We use 64 mel-spaced frequency bins and transform the magnitude using log to arrive at the log-mel spectogram features for the recordings. Following~\cite{hershey2017cnn}, we train VGG network over AudioSet dataset~\cite{gemmeke2017audio} for extracting features from the spectograms.

\noindent\textbf{\wavvec}:
\wavvec models~\cite{baevski2020wav2vec} are transformer based models and are trained in a semi-supervised way on unlabelled raw audios. The models can be finetuned for downstream tasks with task-specific labelled data.  We explore \xlsr model (trained over 53 languages with little overlap with Indic languages), \clsril~\cite{gupta2021clsril} (trained on Indic languages) and \vak \footnote{https://huggingface.co/Harveenchadha/vakyansh-wav2vec2-hindi-him-4200} which was finetuned using 4200 hours of labelled Hindi data over \clsril. For extracting \wavvec features, we pass the raw recordings as input.

\subsection{Model Architecture}
We experiment with \meanpooling, \maxpooling, and recurrent networks (\gru, \lstm). We represent audio recordings by taking average and maximum of the features across temporal dimension in \meanpooling and \maxpooling, respectively. The accumulated features are passed into a fully connected classifier $(512\rightarrow256\rightarrow128\rightarrow2)$ with ReLU activation and 0.1 dropout. For RNNs, audio features are processed through single-layer bidirectional Gated Recurrent Units (\gru) and Long Short-Term Memory (\lstm). The output of the final time step is used as input to the classifier. 
\subsection{Training Setup and Evaluation}
We train the networks using cross entropy loss with Adam optimizer with learning rate of 0.001 and batch size of 16 for 50 epochs. 0.1 is used as dropout for the classification layers. The recordings are normalized by zero-padding shorter audios. We augment data by applying temporal jittering of 0.1 and mask the temporal and frequency/feature dimension randomly between 0 and 10\%.  We report Macro F1 (\fmacro), Accuracy (\accuracy), area under the ROC (\auc) and precision-recall curve (\aucpr) on the test set. 

\section{Results}
\subsection{Monolingual Experiments}
From Table~\ref{table:architectures}, we note that \clsril outperforms other backbones which can be attributed to the pretraining of \clsril in Indic languages. Surprisingly, \vgg which has been trained for identifying different sounds instead of spoken text demonstrates superior performance than \xlsr model for Hindi. However, \vak which has been finetuned for Hindi outperforms \xlsr showing the advantage of language specific finetuning. The RNN baselines overfit during the training and do not improve the results. We evaluate the baselines on other languages in Table~\ref{table:mono_lingual_results} and note that \wavvec models work better for majority of the languages. This can be attributed to the pretraining of these models on speech data. 

\begin{table}[t]
\footnotesize
\caption{Accuracy (\accuracy), Macro F1 (\fmacro), Area under ROC curve (\auc) and Area under Precision-Recall curve (\aucpr) for different architectures and backbones for Hindi.}
\begin{center}
\begin{tabular}{ llcccl } 
\hline
Backbone & Model & \accuracy & \fmacro & \auc &\aucpr \\ [0.5ex] 
\hline
\multirow{4}{*}{\vgg} & \maxpooling  & 78.05 & 78.01  & 0.84 & 0.85 \\
& \meanpooling & 78.59 & 78.57 & 0.85 & 0.86\\
& \lstm  & 77.77 & 77.71  & 0.84 & 0.86 \\
& \gru & 78.57 & 78.56 & 0.85 & 0.86\\
\hline
\multirow{2}{*}{\xlsr} & \maxpooling & 76.96 & 76.90 & 0.84 & 0.84 \\
& \meanpooling & 77.51 & 77.34 & 0.83 & 0.85\\
\hline
\multirow{2}{*}{\vak} & \maxpooling & 79.13 & 79.03 & 0.85 & 0.86\\
& \meanpooling & 78.86 & 78.69 & 0.85 & 0.85\\
\hline
\multirow{4}{*}{\clsril} & \maxpooling &  \textbf{79.67} & \textbf{79.48} & 0.86 & 0.86\\
& \meanpooling & 78.59 & 78.59 & 0.86 & 0.84\\
& \lstm & 70.19 & 69.53 & 0.78 & 0.79 \\
& \gru & 75.34 & 75.23 & 0.82 & 0.83 \\
\hline

\end{tabular}
\end{center}
\label{table:architectures}
\end{table}

% \begin{table}[t]
% \caption{Accuracy (\accuracy) and Macro F1 (\fmacro) score for different architectures and backbones for Hindi.}
% \begin{center}
% \begin{tabular}{ ll|rr } 
% \hline
% Backbone & Model & \accuracy & \fmacro \\ [0.5ex] 
% \hline
% \multirow{2}{*}{\vgg} & \maxpooling  & 78.05 & 78.01 \\
% & \meanpooling & 78.59 & 78.57 \\
% \hline
% \multirow{2}{*}{\xlsr} & \maxpooling & 76.96 & 76.90 \\
% & \meanpooling & 77.51 & 77.34 \\
% \hline
% \multirow{4}{*}{\vak} & \maxpooling & 79.13 & 79.03 \\
% & \meanpooling & 78.86 & 78.69 \\
%  & \gru & 77.50 & 77.50 \\
% & \lstm  & 77.10 & 77.02 \\
% \hline
% \multirow{2}{*}{\clsril} & \maxpooling &  \textbf{79.67} & \textbf{79.48} \\
% & \meanpooling & 78.59 & 78.59 \\
% \hline

% \end{tabular}
% \end{center}

% \label{table:architectures}
% \end{table}

\begin{table}[h]
\footnotesize
\caption{Accuracy (\accuracy) and F1 (\fmacro) for languages for \vgg and \xlsr (\meanpooling) and \clsril features with \maxpooling aggregation.}
\begin{center}
\begin{tabular}{ l|rr|rr|rr } 
\hline
 & \multicolumn{2}{c|}{\vgg} &  \multicolumn{2}{c|}{\xlsr} &  \multicolumn{2}{c}{\clsril}\\ [0.5ex] 
\hline
Lang & \accuracy & \fmacro & \accuracy & \fmacro & \accuracy & \fmacro\\ [0.5ex] 
\hline
Hi & 78.59 & 78.57 & 77.51 & 77.34 & \textbf{79.67} & \textbf{79.48}\\

Be & 78.65 & 77.63 & \textbf{81.08} & \textbf{79.46} & 79.73 & 77.95\\

Pu & 82.01 & 81.97 & 82.01 & 81.99 & \textbf{82.01} & \textbf{82.01}\\

Ha & \textbf{81.15} & \textbf{81.12} & 80.05 & 79.91 & 79.23 & 79.10\\

Ka & 82.91 & 80.06 & \textbf{82.92} & \textbf{80.15}  & 79.67 & 75.39\\
Od & 81.64 & 81.46 & \textbf{83.29} & \textbf{82.21} & 81.64 & 80.21\\
Bh & 76.19 & 71.85 & \textbf{76.48} & 71.10  & 75.89 & \textbf{72.30}\\
Gu & 79.56 & 74.11 & 79.28 & 69.21  & \textbf{80.94} & \textbf{76.38}\\
Ta & 79.78 & 70.77 & \textbf{80.59} & \textbf{75.04}  & 80.59 & 73.39\\
Ma & 81.72 & 77.31 & 81.70 & 75.45  & \textbf{86.29} & \textbf{83.41}\\
\hline
\end{tabular}
\end{center}

\label{table:mono_lingual_results}
\end{table}

\subsection{Cross-lingual Experiments}
In Table~\ref{table:cross_lingual_results}, we train zero-shot models on the source language and evaluate the performance on the target language using \clsril and \maxpooling. The cross-lingual performance is competitive and even better for some languages showing strong cross learning among languages for this task. We hypothesize that models are able to leverage audio properties like pitch, emotions, intensity etc. for this task instead of relying on the actual words, which is highly encouraging. We also combine the data (\all) for all these languages together for training and evaluate on each language separately. We note that combination of all the languages shows improvement for majority of the languages paving path for truly multilingual models for profanity detection.

\begin{table}[h]
\caption{Macro F1 score across languages using \clsril model with \maxpooling aggregation. \textbf{Bold} represent the best combinations.}
\begin{center}
\begin{tabular}{ l|rrrrr } 
\hline
source/target & Hi & Be & Pu & Ka & Ta\\ [0.5ex] 
\hline
Hi & \textbf{79.5 }& 77.0 & 81.5 & \textbf{79.1} & \textbf{74.7}\\
\hline
Be & 78.3 & \textbf{77.9} & 82.0 & 75.2 & 74.1\\
\hline
Pu & 78.5 & 77.1 & 82.0 & 77.6 & 71.0\\
\hline
Ka & 78.3 & 77.4 & 82.4 & 75.4 & 74.1\\
\hline
Ta & 77.1 & 76.0 & \textbf{83.4} & 77.1 & 73.3\\
\hline
\hline
All & 80.7 & 79.1 & 83.4 & 78.4 & 75.2 \\
\hline
\end{tabular}
\end{center}
\label{table:cross_lingual_results}
\end{table}

\section{Conclusion and Future Work}
Detection of abusive content in spoken text is an important problem. Performing ASR followed by a NLP layer for processing the transcription introduces complexity and cost of developing ASR models. In this paper, we contribute a novel and diverse multilingual audio dataset - \prima for tackling this problem entirely in the audio domain. The dataset covers 10 Indic languages with 11,775 samples (65 hours) spoken by 6446 unique users and annotated by expert team of reviewers. We also perform comprehensive experiments and report baselines for encouraging further exploration in this direction.

\noindent \textbf{Ethical Considerations}: Keeping in mind the sensistive nature of the task, we ensure to mandate certain ethical considerations throughout the course of this research and public release of data. Specifically, ShareChat's \textit{public} chatrooms are open for anyone to join and users' informed consent is sought for recording and broadcasting the discussions. Further, we remove any Personally Identifiable Information (PII) from the dataset and anonymize it. The raw data is kept on secure servers with strong access restrictions to prevent any malicious usage.

\vfill\pagebreak

\bibliographystyle{IEEEbib}
\bibliography{strings,refs}
%\input{sections/rebuttal_r1}
%\vfill\pagebreak
%\input{sections/rebuttal_r2}
%\vfill\pagebreak
%\input{sections/rebuttal_r3}
%\vfill\pagebreak
\end{document}